\documentclass[showpacs,preprint,amsmath,amssymb]{revtex4}

\usepackage{graphicx}
\usepackage{dcolumn}
\usepackage{bm}

\begin{document}

\title{
{\bf{Fusion of $^{6}$Li with $^{159}$Tb} at near barrier energies}
}

\author{M. K. Pradhan$^{1}$}
\author{A. Mukherjee$^{1}$}
\email{anjali.mukherjee@saha.ac.in}
\author{P. Basu$^{1}$}
\author{A. Goswami$^{1}$} 
\author{R. Kshetri$^{1}$}
\author{R. Palit$^{2}$}
\author{V. V. Parkar$^{3}$}
\author{M. Ray$^{4}$} 
\author{Subinit Roy$^{1}$}
\author{P. Roy Chowdhury$^{1}$}
\author{M. Saha Sarkar$^{1}$} 
\author{S. Santra$^{3}$}

\affiliation{${^1}$ Nuclear Physics Division, Saha Institute of Nuclear 
Physics, 1/AF, Bidhan Nagar, Kolkata-700064, India }
\affiliation{${^2}$ Department of Nuclear $\&$ Atomic Physics, Tata Institute 
of Fundamental Research, Mumbai-400005, India }
\affiliation{${^3}$ Nuclear Physics Division, Bhabha Atomic Research Centre, 
Mumbai-400085, India }
\affiliation{${^4}$ Department of Physics, Behala College, Parnasree, 
Kolkata-700060, India}


\begin{abstract}

Complete and incomplete fusion cross sections for $^{6}$Li+$^{159}$Tb have 
been measured at energies around the Coulomb barrier by the $\gamma$-ray 
method. The measurements show that the complete fusion cross sections at 
above-barrier energies are suppressed by $\sim$34\% compared to the coupled 
channels calculations. A comparison of the complete fusion cross sections 
at above-barrier energies with the existing data of $^{11,10}$B+$^{159}$Tb and 
$^{7}$Li+$^{159}$Tb shows that the extent of suppression is correlated with 
the $\alpha$-separation energies of the projectiles. It has been argued that 
the Dy isotopes produced in the reaction $^{6}$Li+$^{159}$Tb, at below-barrier 
energies are primarily due to the $d$-transfer to unbound states of $^{159}$Tb,
while both transfer and incomplete fusion processes contribute at 
above-barrier energies.
\end{abstract}


\maketitle
\section{\label{sec:level1}Introduction}
Near barrier fusion is governed by the structure of the interacting nuclei and 
the coupling to the direct nuclear processes, such as inelastic excitation 
and nucleon transfer \cite{mbe88,mdg98}. For nuclear systems with tightly 
bound nuclei, the coupling of the relative motion to these internal degrees 
of freedom successfully explains the enhancement of fusion cross sections 
with respect to the 1-D Barrier Penetration Model (BPM) calculations at 
sub-barrier energies \cite{mdg98}.
However, the situation gets more complicated in reactions involving weakly 
bound nuclei, since they may break up prior to fusion. The interest in 
understanding the influence of breakup on fusion and other reaction processes
has indeed received a fillip in the recent years, especially because of the 
recent advent of the radioactive ion beam facilities in different laboratories 
around the world.

        Owing to the low intensities of the radioactive ion beams currently 
available experimental investigation of reaction mechanisms with unstable 
beams is still difficult, though measurements are being increasingly reported 
\cite{Feko95,Peni95,Fomi95, Yosh96,Rehm98,Kola98,Trot00,Sign04,Sign99}. 
On the contrary, precise fusion cross sections measurements can be carried out 
with the readily available high intensity beams of weakly bound stable nuclei, 
$^{6,7}$Li and $^{9}$Be, which have significant breakup probabilities. 
Such studies with weakly bound stable projectiles may serve to be an important 
step towards the understanding of the influence of breakup on fusion process.

        During the past few years, the effect of breakup of weakly bound 
nuclei on the fusion process has been extensively investigated. 
In fusion measurements of weakly bound stable projectiles with heavy targets
\cite{Sign99,Sign98,Dasg99,Dasg02,Trip02,YWWu03,Dasg04,Trip05,Mukh06,
Goma06,Gomb06,Rath09}, 
events corresponding to the complete fusion (CF) of the projectile with 
the target could be separated experimentally from those resulting due to
the incomplete fusion (ICF) process (where part of the projectile is 
captured by the target). The works show that the CF cross sections are 
substantially suppressed at above barrier energies, compared to the 
predictions of the 1-D BPM calculations. This has been attributed to the 
breakup of the weakly bound projectiles, prior to reaching the fusion barrier.  

By contrast, fusion measurements for medium and light mass systems 
\cite{Mora00,Gome04,Beck03,Mukh96,Mukh98,Mukh99,Mukh00,Mukh02,Mray08}, 
where CF and ICF products could not be experimentally distinguished, only 
total fusion (CF+ICF) cross sections were measured. Such measurements show 
no significant effect of breakup on total fusion at above barrier energies.

     Systematic fusion excitation functions measurement carried out by the 
characteristic $\gamma$-ray method, for the systems $^{10,11}$B+$^{159}$Tb and 
$^{7}$Li+$^{159}$Tb \cite{Mukh06}, shows that the CF cross sections at above 
barrier energies are suppressed for the systems $^{10}$B+$^{159}$Tb and 
$^{7}$Li+$^{159}$Tb  by $\sim$14\% and $\sim$26\% respectively, with respect 
to the coupled channels (CC) calculations. Also, the CF suppression was found 
to be correlated with the $\alpha$-breakup threshold of the projectiles. 
In the context of these results, it appears worthwhile to measure the CF cross 
sections for the system $^{6}$Li+$^{159}$Tb, in view of the fact that $^{6}$Li 
has the lowest $\alpha$-breakup threshold (1.45 MeV) amongst the stable 
projectiles $^{6,7}$Li, $^{9}$Be and $^{10,11}$B. The present work deals with 
the measurement of CF and ICF cross sections for $^{6}$Li+$^{159}$Tb at 
energies around the Coulomb barrier, using the $\gamma$-ray method. To check 
the consistency of the present results with those of Ref. \cite{Mukh06}, the 
reaction $^{7}$Li+$^{159}$Tb was repeated at a few energies in the present 
work. Some preliminary results of the measurement have been reported in a 
conference proceedings \cite{Mukh10}.  

\section{\label{sec:level2}Experimental Details}
\begin{figure}
\vspace{-80pt}
\begin{center}
\includegraphics[height=20cm]{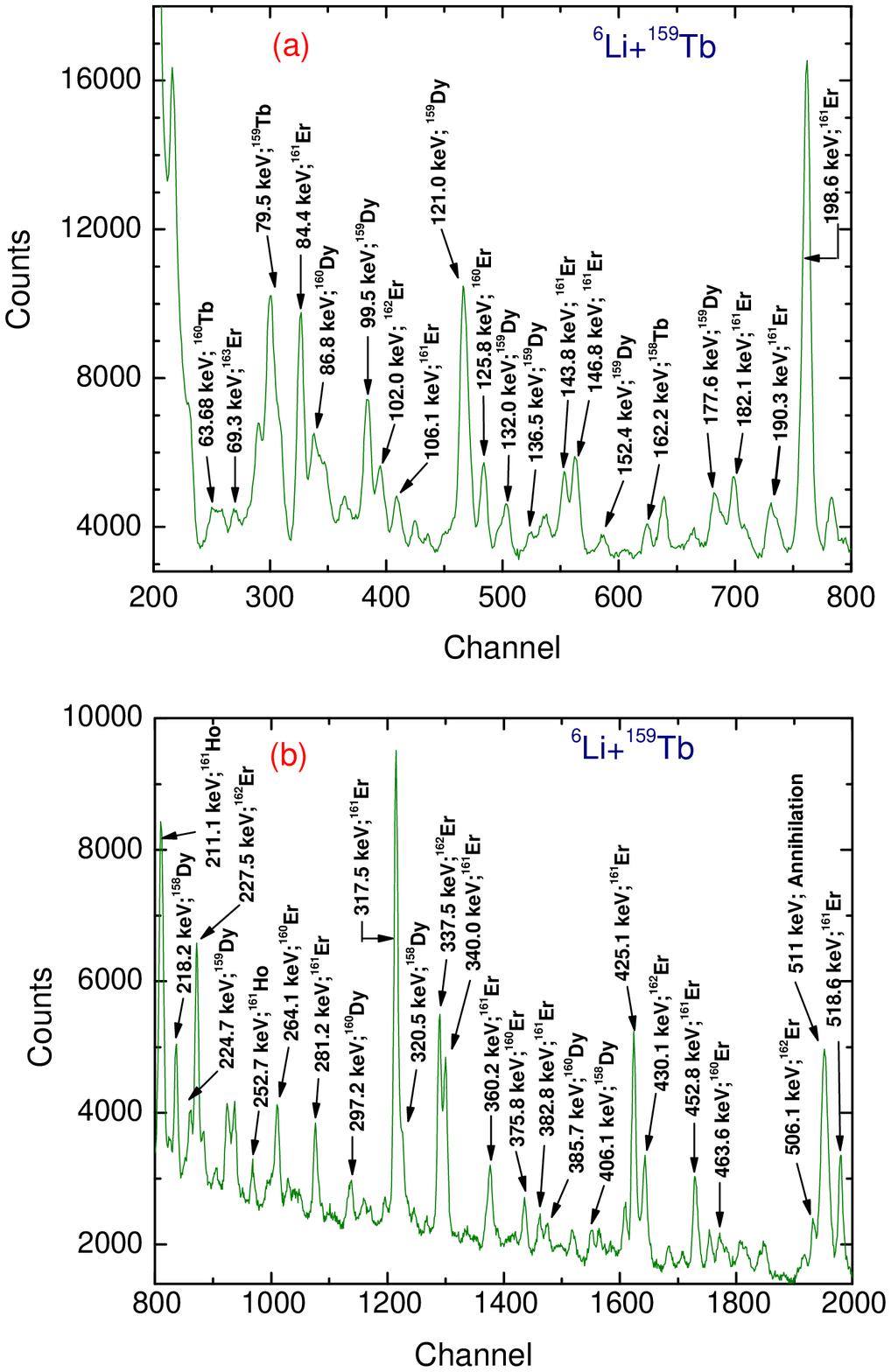}
\end{center}
\vspace{-80pt}
\caption{\label{fig:6Li+159Tb_gamma_spectra} (Color online) Typical 
$\gamma$-ray spectrum obtained with a clover detector placed at 125$^{\circ}$,
for the reaction $^{6}$Li+$^{159}$Tb, at a bombarding energy of 39 MeV.}
\end{figure}

The experiment was performed using the 14UD BARC-TIFR Pelletron accelerator 
at Mumbai. Beams of $^{6}$Li in the energy range 23-39 MeV and $^{7}$Li at 
energies of 28, 34 and 37 MeV bombarded a self-supporting $^{159}$Tb foil of
thickness 1.59$\pm$0.08 mg/cm$^{2}$. To monitor the beam and also for
normalization purposes, two Si-surface barrier detectors were placed at 
$\pm$30$^{\circ}$ about the beam axis inside a spherical reaction chamber of 
22 cm diameter. The total charge of each exposure was measured in a 
1 m long Faraday cup placed after the target. The $\gamma$-rays emitted by the 
reaction products were detected in an absolute efficiency calibrated Compton 
suppressed clover detector, placed at $+125^{\circ}$ with respect to the beam 
direction. An HPGe detector having Be window was placed at $-125^{\circ}$ with 
respect to the beam direction, mainly to detect the low energy gamma rays. Both
online and offline $\gamma$-spectra were taken during the runs, using the Linux
based data acquisition software LAMPS \cite{Lamp09}.   
The absolute efficiencies of the $\gamma$-ray detectors were determined using 
the standard calibrated radioactive sources ($^{152}$Eu, $^{133}$Ba, 
$^{209}$Bi, $^{60}$Co, $^{137}$Cs) placed at the same geometry as the target. 
The target thickness was determined using the 137.5 keV 
(7/2$^{+}$$\rightarrow$3/2$^{+}$(g.s.)) Coulomb excitation 
line of $^{159}$Tb. The same target was used for all the beam exposures. So
to minimize the accumulation of radioactivity in the target, the
target irradiations were carried out from the lowest beam energy onwards. 
A typical $\gamma$-ray addback spectrum from the clover detector, at the 
bombarding energy of 39 MeV, is shown in Figs. 
\ref{fig:6Li+159Tb_gamma_spectra}(a-b). The nuclei produced in the reaction 
were identified by their characteristic $\gamma$-ray energies and are labelled 
in the figure.  

\section{\label{sec:level3}Determination of complete fusion yields}

The compound nuclei $^{165}$Er and $^{166}$Er formed following the CF of 
$^{159}$Tb with $^{6}$Li and $^{7}$Li, respectively, decay predominantly by 
neutron evaporation. This is also predicted by the statistical model 
calculations done using the code PACE \cite{Gavr80}. In the measured energy 
range the evaporation of two to five neutrons occurs, resulting in the 
formation of $^{163-160}$Er and $^{164-161}$Er evaporation residues (ERs) for 
the reactions $^{6}$Li+$^{159}$Tb and $^{7}$Li+$^{159}$Tb, respectively. 

In determining the ER cross sections, the online spectra were mostly used. But 
as and when required, the offline-spectra were also used. It needs to be 
mentioned here that in situations where the ERs are stable, only the in-beam 
$\gamma$-ray spectroscopy method can be used. However, in cases where the 
unstable ERs undergo further radioactive decay to populate the excited states 
of their daughter nuclei, which in turn decay to their ground states by 
emitting $\gamma$-rays, one can also use the off-beam $\gamma$-ray method, if 
the situation is favourable. In the present work, this could be done only for 
the 4{\textit n} channel residual nucleus, $^{161}$Er with a half life 
($T_{1/2}$) of 3.21 hours, produced in the reaction $^{6}$Li+$^{159}$Tb. The 
off-beam $\gamma$-ray method could not be used for the ER $^{163}$Er 
($T_{1/2}$= 75 mins.), as 99.9\% of $^{163}$Er undergoes EC decay to ground 
state of $^{163}$Ho. Also, as the same target was used for all the 
irradiations, the off-beam method could not be used for the ER $^{160}$Er, 
having $T_{1/2}$= 28.58 hours which is substantially large compared to the 
data accumulation times (typically $\sim$1-2 hours).   

While analyzing the data from the clover detector, the addback spectra were 
used. Wherever possible, the cross sections obtained from the clover detector 
spectra were compared with those from the HPGe detector spectra, and they 
were found to be in good agreement.\\ 

The $\gamma$-ray cross sections ($\sigma_{\gamma}$) were obtained from the 
relation
\begin{equation}
\sigma_{\gamma}=\frac{N_{\gamma}}{(\epsilon_{\gamma}N_{B}N_{T})}
\end{equation}
where N$_{\gamma}$ is the number of counts under the $\gamma$-ray peak,
$\epsilon_{\gamma}$ is the absolute full energy peak detection efficiency
of the detector for the specific $\gamma$-ray, N$_{B}$ is the total number
of beam particles incident on the target and N$_{T}$ is the number of target 
nuclei per cm$^2$. The quantity N$_{B}$ was determined by dividing the 
charge Q collected in the Faraday cup by the equilibrium charge value 
$\bar{Z}$e, obtained from Ref.\cite{Marion}. 
The total systematic uncertainty in the $\gamma$-ray cross sections, arising 
because of the uncertainties in N$_{B}$, N$_{T}$ and $\epsilon_{\gamma}$, 
is $\sim$8$\%$. This is added in quadrature to the statistical error in 
N$_{\gamma}$ to get the total error in $\sigma_{\gamma}$.

\begin{figure}{}
\vspace{-180pt}
\begin{center}
\includegraphics[height=20cm]{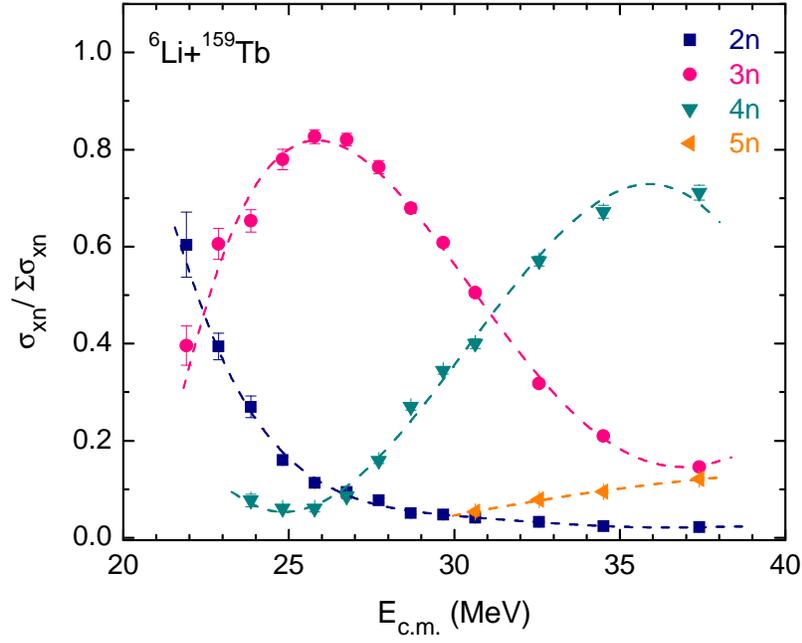}
\end{center}
\vspace{-180pt}
\caption{\label {fig:6Li+159Tb_Frac_chan} (Color online) Ratio of individual 
channel cross sections to the total channel cross sections as a function of the 
centre-of-mass energy for the reaction $^{6}$Li+$^{159}$Tb. The errors are
statistical only. The dashed lines are drawn to guide the eye.}
\end{figure}  

\begin{figure}{} 
\vspace{-180pt}
\begin{center}
\includegraphics[height=20cm]{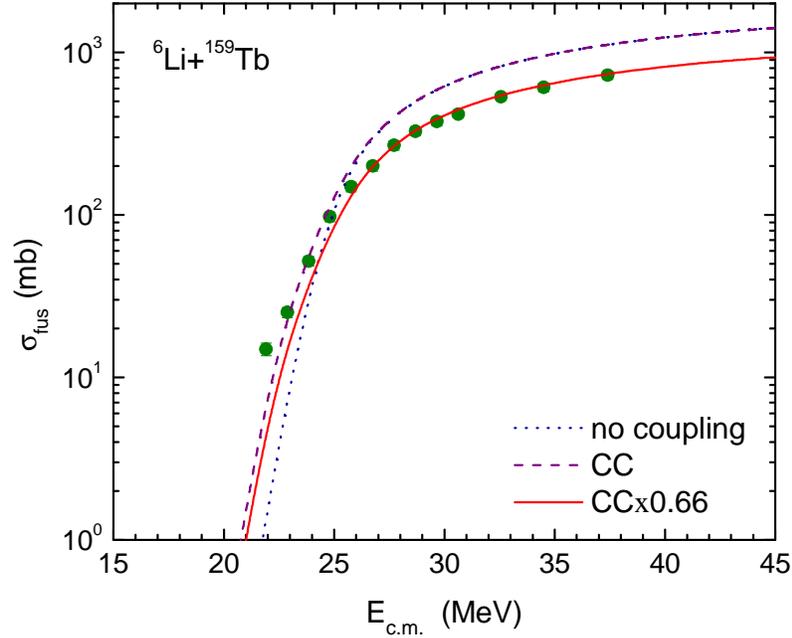}
\end{center}
\vspace{-180pt}
\caption{\label {fig:6Li+159Tb_CF} (Color online) Complete fusion cross 
sections as a function of the centre-of-mass energy for the reaction 
$^{6}$Li+$^{159}$Tb. The error bars indicate the total errors. The dotted and 
dashed lines show the uncoupled and coupled channel calculations, respectively,
performed with the code CCFULL. The solid line is the coupled channels 
calculation multiplied by the factor of 0.66.}
\end{figure}

For the even-even ERs ($^{164,162,160}$Er), the cross sections were extracted 
from the extrapolated value of the intensity at J=0 obtained from the 
measured $\gamma$-ray intensities (after correcting for the internal 
conversion) for various transitions in the ground state rotational band
\cite{Mukh06}. 
For the odd-mass ERs ($^{163,161}$Er) the cross sections were obtained by
adding the cross sections of the $\gamma$-rays corresponding to the 
transitions from the excited states to the ground states of the nuclei, as 
done by Broda {\em{et al.}} \cite{Broda75}. In such cases, however, direct 
population of the  ground states of the nuclei could not be considered. 
Nevertheless, the direct feedings to the ground states are expected to be 
substantially small in this mass and energy region, except at very low 
bombarding energies. In fact, in the present work this has been checked for 
the ER $^{161}$Er, produced in the reaction $^{6}$Li+$^{159}$Tb, as both 
in-beam and off-beam $\gamma$-ray method could be applied to measure its 
production cross sections at low bombarding energies. 
It was observed that the cross sections, obtained from the in-beam 
$\gamma$-rays of $^{161}$Er (where direct population of the ground state is not 
included) and those from the off-beam $\gamma$-rays of $^{161}$Ho,
following EC decay of $^{161}$Er, (which obviously includes direct ground 
state population of $^{161}$Er) are practically same. This shows the ground 
state contribution to be rather small and can safely be ignored in the 
evaluation of the CF cross sections. The CF cross sections for both reactions 
were obtained from the sum of the 2$\textit n$$-$5$\textit n$ ER cross 
sections. 
 
Figure \ref{fig:6Li+159Tb_Frac_chan} shows the individual $\textit xn$ channel 
cross sections normalized to the CF cross sections (fractional channel cross 
sections) for the reaction $^{6}$Li+$^{159}$Tb. The measured CF cross sections, 
along with the total errors, for the reaction $^{6}$Li+$^{159}$Tb are plotted 
in Fig. \ref{fig:6Li+159Tb_CF}. The CF cross sections for $^{7}$Li+$^{159}$Tb,
measured at a few bombarding energies in the same setup, are 
seen to agree well with the earlier measurements \cite{Mukh06,Broda75}, thus 
enabling a reliable comparison of the present results with the earlier ones.

\section {\label{sec:level4}Coupled channels calculations}

To interpret the measured fusion excitation function in a theoretical 
framework, the realistic coupled channels (CC) code CCFULL \cite{Hagi99} was 
used to calculate the fusion cross sections for $^{6}$Li+$^{159}$Tb. The 
initial input potential parameters ($V_{0}, r_{0},$ and $a$) were obtained 
from the Woods-Saxon parametrization of the Aky$\ddot {u}$z-Winther (AW) 
potential \cite{Broglia81}, and are shown in Table I. The table also shows the 
corresponding uncoupled fusion barrier parameters ($V_{b}, R_{b}$, and 
$\hbar {\omega}$). As CCFULL cannot handle shallow potential, a deeper 
potential was used. This modified potential was derived keeping the diffuseness
parameter fixed at $a$=0.85 fm, following the systematic trend of high 
diffuseness required to fit the high energy part of the fusion excitation 
functions \cite{Newton04}. To obtain the appropriate potential, the parameters 
$V_{0}$ and $r_{0}$ were varied accordingly so that the corresponding 1-D BPM
cross sections agree with those obtained using the AW potential parameters at 
higher energies \cite{Mukh06}. The modified potential used for the CC 
calculations, and the corresponding uncoupled barrier parameters are given in 
Table I. Using the modified potential parameters, the 1-D BPM calculations 
were done using the code CCFULL, in the no coupling limit and the results
are shown by the dotted line in Fig. \ref{fig:6Li+159Tb_CF}. The CF 
cross sections at below-barrier energies are seen to be enhanced and the
cross sections at above-barrier energies are found to be reduced compared
to the 1-D BPM calculations. The enhancement at below-barrier energies
may be because of the fact that the target $^{159}$Tb is a well deformed
nucleus. 

The effect of target deformation on the fusion cross sections was calculated 
by including coupling to the ground state rotational band of the target 
nucleus. As described in Ref. \cite{Mukh06}, for the odd-A nucleus $^{159}$Tb,
the excitation energies and deformation parameters were taken to be the 
averages of those of the neighbouring even-even nuclei $^{158}$Gd and 
$^{160}$Dy. The energy states, in the ground state rotational band 
of the corresponding average spectrum ($\beta_{2}$=0.344 \cite{Rama87} and 
$\beta_{4}$=+0.062 \cite{Moll95}), upto 12$^{+}$ were included in the 
calculations. Projectile excitation was not included in the calculations. 
It needs to be mentioned here that $^{6}$Li has a ground state with non-zero 
spin (1$^{+}$) and spectroscopic quadrupole moment of $-$0.082 fm$^{2}$, and 
has an unbound first excited state (3$^{+}$) at 2.186 MeV. But coupling to 
the unbound first excited state of $^{6}$Li with such ground state properties,
along with the rotational coupling to the target excited states could not 
be included in the CCFULL calculations.

The dashed line in Fig. \ref{fig:6Li+159Tb_CF} shows the CC calculations that 
include rotational coupling to the inelastic states of the target. The 
calculations, though reproduce the low energy part of the data reasonably 
well, overestimate the high energy part of the data. The little difference 
that can be seen at the lowest energy could be due to the projectile effect, 
which could not be considered in the calculations, as mentioned. At above 
barrier energies, where coupling is not expected to play any significant role, 
the CF cross sections are found to be suppressed compared to the CC 
calculations. 

As CC model cannot yet separate CF and ICF, the measured CF cross sections can 
only be compared with the calculated total fusion cross sections. So in order
to have an estimate of the extent of CF suppression compared to the total 
fusion cross sections, the CC calculations for $^{6}$Li+$^{159}$Tb were scaled 
so as to reproduce the high energy part of the measured CF excitation function. 
Agreement could be achieved only if the calculated fusion cross sections are
scaled by a factor of 0.66, and the resulting scaled calculations 
are shown in Fig. \ref{fig:6Li+159Tb_CF} by the solid line. The CF suppression 
factor (F$_{CF}$) for the system is thus 0.66$\pm$0.05, where the uncertainty 
of $\pm$5\% has been estimated, resulting from the overall errors 
in the measured fusion cross sections. The CF suppression of 34$\pm$5\% 
thereby obtained at above barrier energies for $^{6}$Li+$^{159}$Tb agrees 
with the value reported for the heavier systems $^{6}$Li+$^{209}$Bi
\cite{Dasg02} and $^{6}$Li+$^{208}$Pb\cite{YWWu03} and is also in close 
agreement with the suppression of 32$\pm$5\% reported for 
$^{6}$Li+$^{144}$Sm \cite{Rath09}.

\begin{table*}
\caption{\label{tab:table1} The parameters for AW and modified CC potentials, 
along with the corresponding derived uncoupled barrier parameters 
V$_{b}$, R$_{b}$, and $\hbar$$\omega$.}
\begin{ruledtabular}
\begin{tabular}{cccccccc}
System & Potential & V$_{0}$ & r$_{0}$ & $\textit a$ & V$_{b}$ & R$_{b}$ & $\hbar$$\omega$ \\
   &    & (MeV)  & ($\textit fm$)   & ($\textit fm$)       & (MeV)  & ($\textit fm$)   & (MeV) \\
\hline
\\
 $^{6}$Li+$^{159}$Tb & AW & 46.40 & 1.18 & 0.62 & 24.89 & 10.60 & 4.85 \\

               & CC & 128.0 & 0.98 & 0.85 & 24.48 & 10.53 & 4.15  \\

 $^{7}$Li+$^{159}$Tb & AW & 46.43 & 1.18 & 0.62 & 24.70 & 10.69 & 4.48 \\

 $^{10}$B+$^{159}$Tb & AW & 54.54 & 1.18 & 0.64 & 40.71 & 10.79 & 4.68 \\

 $^{11}$B+$^{159}$Tb & AW & 54.54 & 1.18 & 0.64 & 40.34 & 10.89 & 4.42 \\

\end{tabular}
\end{ruledtabular}
\end{table*}

\section{\label{sec:level5}Comparison of suppression with other systems}
\begin{figure}{}
\vspace{-130pt}
\begin{center}
\includegraphics[height=20cm]{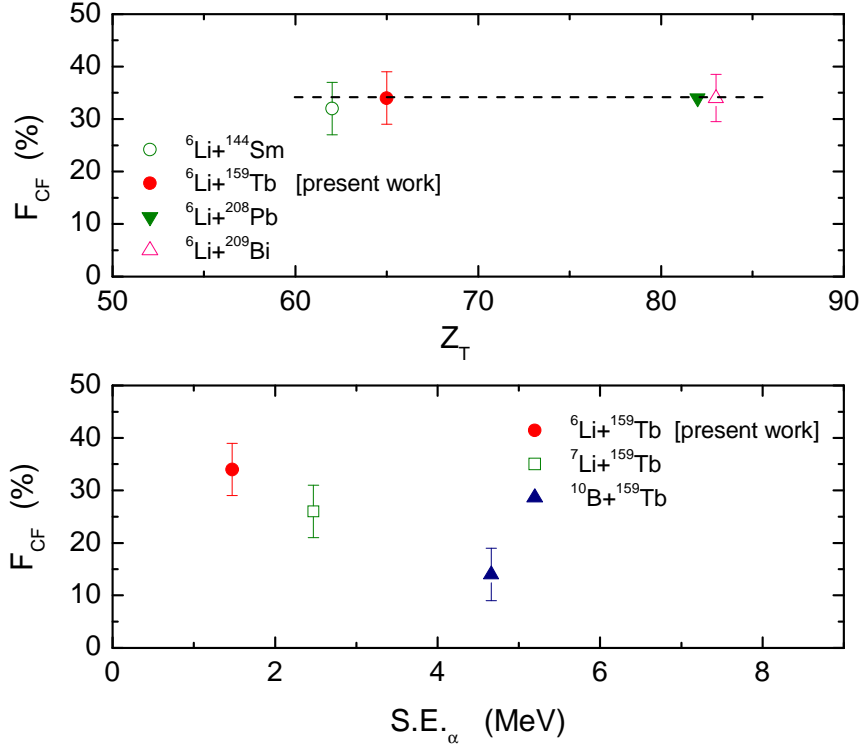}
\end{center}
\vspace{-150pt}
\caption{\label {fig:CF_suppression_ZT_SE} (Color online) (a) CF suppression 
($\%$) as a function of atomic no. $Z_{T}$ of target for the $^{6}$Li-induced 
reactions involving different targets. The reactions considered are $^{6}$Li
incident on $^{144}$Sm \cite{Rath09}, $^{159}$Tb (present work), $^{208}$Pb
\cite{YWWu03}, and $^{209}$Bi \cite{Dasg02}. The dotted line is drawn to guide 
the eye. (b) CF suppression(\%) as a function of $\alpha$-separation energies 
(S.E.$_{\alpha}$) of the projectiles in reactions with the target $^{159}$Tb.
The reactions considered are $^{10}$B+$^{159}$Tb \cite{Mukh06}, 
$^{7}$Li+$^{159}$Tb \cite{Mukh06} and $^{6}$Li+$^{159}$Tb (present work).}
\end{figure}

The F$_{CF}$ for $^{6}$Li induced reactions on different targets 
are compared in Fig. \ref{fig:CF_suppression_ZT_SE}(a), using the present data 
and those reported in the literature \cite{Dasg02, YWWu03, Rath09}. 
The dotted line has been drawn in the figure only to guide the eye. It appears 
that the F$_{CF}$ for $^{6}$Li induced reactions are almost independent of the 
atomic number ($Z_{T}$) of the target nucleus, in the heavy mass region. 
However, more values of F$_{CF}$ for $^{6}$Li induced reactions, especially 
with targets of lower Z$_{T}$ are required before drawing any definite 
conclusion. Figure \ref{fig:CF_suppression_ZT_SE}(b) compares F$_{CF}$ for
the reactions $^{6}$Li+$^{159}$Tb, $^{7}$Li+$^{159}$Tb \cite{Mukh06} and 
$^{10}$B+$^{159}$Tb \cite{Mukh06} as a function of the $\alpha$-separation
energies (S.E.$_{\alpha}$) of the projectiles. Like $^{6}$Li+$^{159}$Tb, a
$\pm$5\% uncertainty has also been estimated for F$_{CF}$ of 
$^{7}$Li+$^{159}$Tb and $^{10}$B+$^{159}$Tb reactions. The plot shows that 
there is a correlation between F$_{CF}$ and S.E.$_\alpha$. But more such 
measurements, including reactions with unstable projectiles, are needed to 
understand the nature of the correlation. 

\begin{figure}{}
\vspace{-140pt}
\begin{center}
\includegraphics[height=20cm]{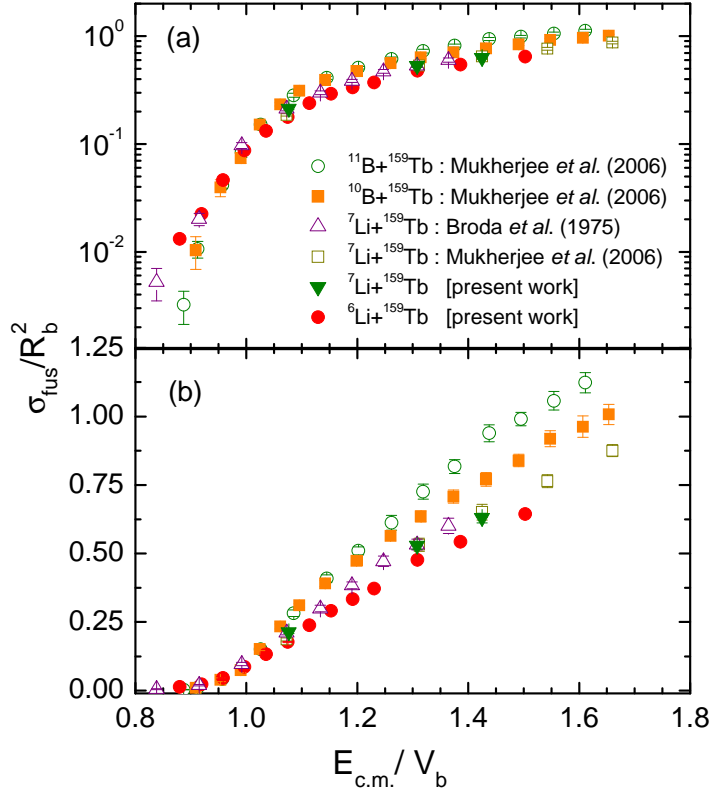} 
\end{center}
\vspace{-140pt}
\caption{\label {fig:red_scale_log_linear} (Color online) A comparison of the 
reduced complete fusion excitation functions for the systems 
$^{10,11}$B+$^{159}$Tb \cite {Mukh06} and $^{7}$Li+$^{159}$Tb \cite{Broda75,
Mukh06} with those of the present measurements for $^{6,7}$Li+$^{159}$Tb. 
The errors are statistical only.}
\end{figure}

Figure \ref{fig:red_scale_log_linear} compares the reduced fusion cross 
sections $\sigma_{fus}/$R$_{b}^{2}$ as a function of E$_{c.m.}/$V$_{b}$ for 
different projectiles in logarithmic scale (a) and linear scale (b). 
The parameters V$_{b}$ and R$_{b}$ used for the reduction are those deduced 
from the AW potentials, and are listed in Table I. The CF cross sections for 
$^{11,10}$B+$^{159}$Tb and $^{7}$Li+$^{159}$Tb were obtained from 
Refs.\cite{Mukh06,Broda75}. It can be seen from Fig.
\ref{fig:red_scale_log_linear}(a), that at the lowest energies the CF cross 
sections of $^{6,7}$Li+$^{159}$Tb are enhanced compared to those of 
$^{10,11}$B+$^{159}$Tb reactions.  This enhancement, which could be due to the 
effect of the projectiles $^{6,7}$Li, was also observed while comparing the 
measurements with CCFULL calculations [Fig. \ref{fig:6Li+159Tb_CF} and 
Ref.\cite{Mukh06}]. For the reaction $^{6}$Li+$^{159}$Tb, this has already been
discussed in Sec. IV. For the reaction $^{7}$Li+$^{159}$Tb, the deformation of 
$^{7}$Li needs to be considered in the calculations \cite{Mukh06}, but both 
projectile and target deformations can not be included simultaneously in the 
CCFULL calculations.

Figure \ref{fig:red_scale_log_linear}(b) shows that as one moves from the 
projectile $^{11}$B to $^{6}$Li, {\em{i.e.}} as the projectile $\alpha$-breakup 
threshold decreases, the CF cross sections are 
observed to be more and more suppressed. A comparison with the CCFULL 
calculations has shown that the measured CF cross sections for 
$^{10}$B+$^{159}$Tb, $^{7}$Li+$^{159}$Tb \cite{Mukh06} and $^{6}$Li+$^{159}$Tb 
are suppressed by $\sim$14\%, $\sim$26\% and $\sim$34\% respectively. This 
certainly shows that the CF suprression is correlated with the $\alpha$-breakup
threshold of the projectile. Lower the $\alpha$-breakup threshold, larger is 
the CF suppression. Thus the CF suppression can be attributed to the loss of 
flux from the fusion channel due to the breakup of the loosely bound 
projectiles, and hence at least a major part of this suppression should be the 
ICF cross sections of the reactions. Also, if one looks carefully into Fig. 
\ref{fig:red_scale_log_linear}(b), it appears that higher the $\alpha$-breakup 
threshold of the projectile, higher is the energy where the CF suppression 
starts. However, more such systematic measurements, especially with unstable 
beams, are required for confirming this observation.

\section{\label{sec:level6}Incomplete fusion}

In order to have a complete picture of the fusion process in the reaction
$^{6}$Li+$^{159}$Tb, besides CF cross sections, it is also important to 
measure the ICF cross sections. As discussed in the previous section, a 
major part of the observed reduction in CF is expected to be due to the 
ICF process. 

\begin{figure}{}
\vspace{-150pt}
\begin{center}
\includegraphics[height=20cm]{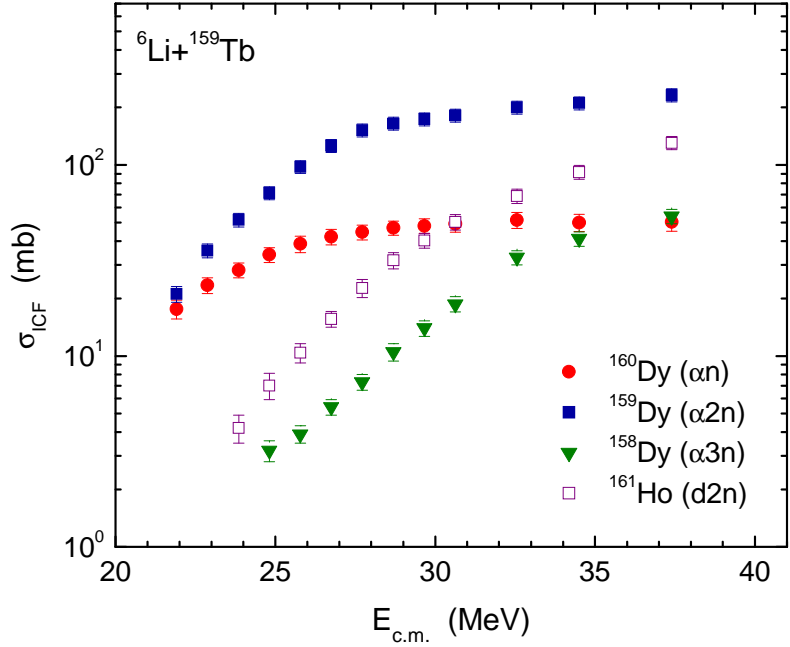}
\vspace{-180pt}
\end{center}
\caption{\label {fig:6Li+159Tb_ICF} (Color online) The ICF/transfer cross 
sections measured for the reaction $^{6}$Li+$^{159}$Tb. The cross sections 
corresponding to the $\alpha$$\textit n$, $\alpha$2$\textit n$ and 
$\alpha$3$\textit n$ channels, following $d$-capture by the target, and the 
cross sections corresponding to the $d2n$ channel, following $\alpha$-capture 
by the target are shown.} 
\end{figure}

In the $\gamma$-ray spectra, besides the $\gamma$-ray lines of the Er nuclei 
resulting from CF, the $\gamma$-ray lines corresponding to Dy and Ho isotopes 
produced via the ICF processes were also observed. In the reaction 
$^{6}$Li+$^{159}$Tb, the Dy nuclei are produced by the capture of the lighter 
projectile fragment, $\textit d$, following $^{6}$Li breakup, by the target 
$^{159}$Tb and subsequent emission of neutrons. Similarly, the Ho nuclei are 
formed by the capture of the heavier projectile fragment, $\alpha$,  by 
$^{159}$Tb, followed by neutron emission. The ICF cross sections are shown in 
Fig. \ref{fig:6Li+159Tb_ICF}. The cross sections of the ICF products were 
determined in a similar way as that for the CF residues. The 
$\alpha$$\textit n$, $\alpha$2$\textit n$ and $\alpha$3$\textit n$ channels, 
following the capture of $\textit d$ by $^{159}$Tb, are seen to be the dominant
ICF channels. On the other hand, only $\gamma$-lines corresponding to 
$^{161}$Ho nucleus resulting from the $\alpha$+$^{159}$Tb ICF process, 
followed by 2$\textit n$ emission, could be identified in the spectra. 
However, the ICF contribution of $^{161}$Ho, plotted in the figure, partly
includes the contribution of $^{161}$Ho produced via the EC decay
of $^{161}$Er CF residue. Nevertheless, it is clear that the  contribution of 
$^{161}$Ho formed in the ICF process is relatively much less compared to Dy 
isotopes. A possible explanation of this could be given on the basis of 
Q-values of the reactions. It is to be noted that the Q-value for the reaction 
$^{159}$Tb($^{6}$Li,$\alpha$)$^{161}$Dy is $+10.2$ MeV, while it is $-2.2$ MeV 
for the reaction $^{159}$Tb($^{6}$Li,$\textit d$)$^{163}$Ho. This indicates 
that the former channel corresponding to ICF process, where the 
$\alpha$-particle is emitted with the $\textit d$ being captured by the target 
is more favoured compared to the latter. Our measurement on the systems 
$^{7}$Li+$^{159}$Tb and $^{10}$B+$^{159}$Tb reported earlier \cite{Mukh06} 
also showed similar result.

It needs to be mentioned here that the ICF cross sections for Dy isotopes 
also include contributions from transfer of $\textit d$ from projectile 
$^{6}$Li to the higher excited states of the target since in the present 
$\gamma$-ray measurement it was not possible to distinguish between the two 
events. Also, the single-proton stripping reaction 
$^{159}$Tb($^{6}$Li,$^{5}$He)$^{160}$Dy, with Q-value +2.836 MeV, if occurs 
will lead to the same $^{160}$Dy nucleus. Hence the contribution from 
$^{160}$Dy nuclei via $\textit p$-transfer, if any, is also included in the 
$\alpha$$\textit n$ channel cross section. 

A careful insight into Fig. \ref{fig:6Li+159Tb_ICF} shows appreciable 
cross sections for Dy nuclei, even at energies below the barrier where CF
shows no suppression (Fig. \ref{fig:6Li+159Tb_CF}). 
This is perhaps because of the fact that at below-barrier energies, it is 
essentially the transfer of $d$ to the unbound states of $^{159}$Tb (one-step 
process), followed by the emission of neutrons, that produces the Dy isotopes. 
In a simplistic picture, this can be understood by considering the optimum 
Q-value (Q$_{opt}$) associated with a transfer reaction. The ground state 
Q-value (Q$_{gg}$) for the $d$-transfer reaction 
$^{159}$Tb($^{6}$Li,$\alpha$)$^{161}$Dy is +10.2 MeV, and Q$_{opt}$ for the 
transfer process, say at $E_{c.m.}$= 22 MeV and 25 MeV are calculated 
\cite{Morgen86} to be
$-$7.1 MeV and $-$8.1 MeV, respectively. The excitation energy ($\epsilon^{*}$)
in $^{161}$Dy to which the $d$-transfer is energetically favoured is given by 
Q$_{gg}-$Q$_{opt}$. Thus at $E_{c.m.}$= 22 MeV and 25 MeV, $\epsilon^{*}$= 
17.3 MeV and 18.3 MeV, respectively, thereby showing that the $d$-transfer to
$^{159}$Tb will energetically favour the production of $^{161}$Dy nuclei in the
unbound states. Unlike transfer, at below-barrier energies, the breakup 
fragments may not have sufficient energy to overcome the Coulomb barrier and 
get further captured by the target (two-step process). By contrast, at 
above-barrier energies the breakup fragments will have sufficient energy to 
undergo further fusion with the target and hence at such energies the ICF 
(breakup-fusion) process, along with $d$-transfer, lead to the production of Dy 
nuclei. It is mainly the ICF (breakup-fusion) yield (which could not be 
separated from transfer in the present measurement) that contributes to the 
reduction of CF at above barrier energies. Similar argument also holds true
for the Ho nuclei. Unfortunately, only one Ho isotope, namely $^{161}$Ho could 
be identified in the present work and that too had an admixture due to the 
contribution from $^{161}$Ho nuclei resulting from the EC decay of $^{161}$Er
residue. So nothing conclusive could be said about Ho nuclei. Detailed 
exclusive measurements aimed at disentangling ICF and transfer yield, though 
difficult, are indeed necessary to see how much of the reduction in CF is 
accounted for by the ICF process.

\section{\label{sec:level7}Summary}

The CF cross sections for the reaction $^{6}$Li+$^{159}$Tb have been measured
at energies around the Coulomb barrier, using the $\gamma$-ray method. CC
calculations using the code CCFULL were done to calculate the total fusion
cross sections. The calculated fusion cross sections had to be scaled by a 
factor of 0.66$\pm$0.05 to reproduce the measured CF cross sections at above 
barrier energies. The above-barrier CF suppression has been attributed to the 
breakup of the loosely bound $^{6}$Li nucleus. The CF suppression of 
$\sim$34$\%$ for $^{6}$Li+$^{159}$Tb when compared to the values of 
$\sim$26$\%$ and $\sim$14$\%$ for $^{7}$Li+$^{159}$Tb and $^{10}$B+$^{159}$Tb 
\cite{Mukh06} respectively, convincingly shows that the CF suppression is 
correlated with the $\alpha$-separation energy of the projectile. Lower the 
$\alpha$- breakup threshold of the projectile, larger is the CF suppression. 
At energies below the barrier, enhancement of CF cross sections could be 
reasonably well reproduced by considering the deformation of the target.  

The nuclei produced via the ICF process in the reaction $^{6}$Li+$^{159}$Tb
were also identified and their cross sections have been determined. Similar to 
$^{10}$B+$^{159}$Tb and $^{7}$Li+$^{159}$Tb \cite{Mukh06}, the present 
measurement also shows that the $\alpha$-emitting channel is the favoured ICF 
process in reactions of projectiles, having low $\alpha$-breakup thresholds, 
with $^{159}$Tb target.

At below barrier energies, the Dy isotopes are primarily produced by the 
$d$-transfer to the unbound states of $^{159}$Tb, while at above barrier
energies both transfer and ICF processes contribute to their production.

Further investigation of the light particles emitted in reactions 
involving loosely bound projectiles, in conjunction with the results presented 
here, may lead to a better understanding of the mechanisms involved in such 
reactions.  

\begin{acknowledgments}
We are grateful to Prof. B.K. Dasmahapatra for valuable discussions and advices
at various stages of the work. We thank Mr. P.K. Das for his earnest technical 
help during the experiment. We would also like to thank the accelerator staff 
at the BARC-TIFR Pelletron Facility, Mumbai, for their untiring efforts in
delivering the beams.
\end{acknowledgments}

\end{document}